# Unconventional Planar Hall Effect in Exchange-Coupled Topological Insulator-Ferromagnetic Insulator Heterostructures

David Rakhmilevich[1,3], Fei Wang[2], Weiwei Zhao[2], Moses H. W. Chan[2], Jagadeesh S. Moodera[1,4], Chaoxing Liu[2], and Cui-Zu Chang[1,2*]

[1]Francis Bitter Magnet Lab, Massachusetts Institute of Technology, Cambridge, MA 02139, USA

[2]Department of Physics, The Pennsylvania State University, University Park, PA 16802, USA

[3]School of Engineering and Applied Sciences, Harvard University, Cambridge, MA 02139, USA

[4]Department of Physics, Massachusetts Institute of Technology, Cambridge, MA 02139, USA

Corresponding author: cxc955@psu.edu (C.Z.C.)

The Dirac electrons occupying the surface states (SSs) of topological insulators (TIs) have been predicted to exhibit many exciting magneto-transport phenomena. Here we report on the first experimental observation of an unconventional planar Hall effect (PHE) and an electrically gate-tunable hysteretic planar magnetoresistance (PMR) in EuS/TI heterostructures, in which EuS is a ferromagnetic insulator (FMI) with an in-plane magnetization. In such exchange-coupled FMI/TI heterostructures, we find a significant (suppressed) PHE when the in-plane magnetic field is parallel (perpendicular) to the electric current. This **behavior differs from previous observations of the PHE in ferromagnets and semiconductors. Furthermore, as the thickness of the 3D TI films is reduced into the 2D limit, in which the Dirac SSs develop a hybridization gap, we find a suppression of the PHE around the charge neutral point indicating the vital role of Dirac SSs in this phenomenon. To explain our findings, we outline a symmetry argument that excludes linear-Hall mechanisms and suggest two possible non-linear Hall mechanisms that can account for all the essential qualitative features in our observations.**

The Hall effect or the appearance of a voltage transverse to an electric current for electric conductors placed in an external magnetic field is among the most well-known magneto-transport phenomena [1]. The ordinary Hall effect, arising from the Lorentz force experienced by current carriers, requires the magnetic field to be perpendicular to both the electric current direction and the sample plane. However, a transverse voltage can also emerge in certain systems when the magnetic field is in the plane of the sample and electric current, a phenomenon known as the planar Hall effect (PHE). The PHE was experimentally observed in bulk ferromagnets [2], nano-crystalline $Co_{60}Fe_{20}B_{20}$ [3], magnetic semiconductors such as (Ga,Mn)As [4], nonmagnetic semiconductors like germanium [5] and topological insulator (TI) films [6]. The PHE cannot be induced by the Lorentz force, and various microscopic mechanisms have been proposed for this phenomenon, including anisotropic scattering by impurities [2,6], a non-spherical Fermi surface [5], spin Hall magnetoresistance [7] and chiral anomaly [8-10]. In particular, it has been shown that the PHE is usually related to anisotropic magnetoresistance (AMR) and in the absence of spontaneous resistivity anisotropy in the crystal, both effects could be described by the following phenomenological equations [11,12]:

$$\rho_{PHR} = (\rho_{//} - \rho_{\perp})\sin\Phi\cos\Phi \tag{1}$$

$$\rho_{AMR} = \rho_{\perp}(1 + \frac{\rho_{//} - \rho_{\perp}}{\rho_{\perp}}\cos^2\Phi) \qquad (2)$$

where $\rho_{PHR}$ and $\rho_{AMR}$ are the transverse and longitudinal magnetoresistances, respectively, $\Phi$ is the angle between the current *I* and the in-plane magnetic field *B*, while $\rho_{//}$ ($\rho_{\perp}$) is the resistance of the sample when *I* is parallel (perpendicular) to *B*. The above angular dependence was shown to be consistent with recent experimental studies on metallic and semiconducting ferromagnets [3,4] as well as on non-magnetic TIs under high magnetic fields [6]. Specifically, according to Eq. (1) the planar Hall resistance (PHR) is zero when *B* is parallel ($\Phi = 0º$) or perpendicular ($\Phi = 90º$) to *I* while it is maximized when $\Phi = 45º$.

In this *Rapid Communication*, we present in-plane magneto-transport measurements in EuS/(Bi$_{0.22}$Sb$_{0.78}$)$_2$Te$_3$ heterostructures, as a prototype of ferromagnetic insulator (FMI)/TI heterostructures. Our measurements demonstrate the first observation of a PHE and tunable planar magnetoresistance (PMR) in FMI/TI heterostructures. The gate voltage ($V_g$) dependence of the PHE shows a peak of PHR when the chemical potential is near the Dirac point. Moreover, as the TI film thickness in the heterostructures is reduced from 4 quintuple layers (QL~1nm) to 3QL, in which the Dirac SSs develop a hybridization gap [13-15], the PHE and PMR properties change dramatically indicating the vital role of Dirac SSs in magneto-transport. Interestingly, we observed a significant PHR for in-plane magnetic field *B* parallel to the current *I* while it was suppressed for *B* perpendicular to *I*, in contrast to the conventional PHE as described by Eq. (1). Based on these findings and a simple symmetry argument, we exclude linear-Hall mechanisms and suggest two possible non-linear Hall mechanisms, which capture all the essential qualitative features in our observations.

The EuS/(Bi$_{0.22}$Sb$_{0.78}$)$_2$Te$_3$ heterostructures were grown on 0.25mm thick heat-treated SrTiO$_3$ (111) substrates in a custom-built molecular beam epitaxy (MBE) chamber [16,17]. The Bi:Sb ratio was controlled to locate the chemical potential close to the Dirac point in order to enhance the contribution of the SSs in magneto-transport [16,18-20]. The growth was monitored using *in-situ* reflective high-energy electron diffraction (RHEED). A 5nm EuS (111) layer was deposited over the TI film at room temperature followed by a 4nm thick $Al_2O_3$ capping layer. A representative X-ray diffraction pattern of the 5nm EuS/4QL $(Bi_{0.22}Sb_{0.78})_2Te_3$ heterostructure is shown in **Fig. 1a**. We have previously demonstrated that EuS forms a continuous film when grown on top of $Sb_2Te_3$ films with no inter-diffusion and exhibits a well-defined in-plane magnetization [21]. The transport measurements were carried out in a Hall-bar geometry using conventional direct current techniques (**Fig. 1b** and Supporting Materials) at a base temperature of $T$ = 1K.

A major challenge in transport studies of TIs is to distinguish between the contributions of bulk carriers and SSs. In order to respond to this challenge, we focus on 4QL $(Bi_{0.22}Sb_{0.78})_2Te_3$ films. At such a thickness, the bulk carrier contribution is minimized while preserving the gapless Dirac SSs on the surface [13-15]. In addition, the large dielectric constant of $SrTiO_3$ (111) substrates at low temperatures makes it possible to efficiently tune the chemical potential of the TI films by changing $V_g$ [16,17]. The sheet longitudinal resistance $R_{xx}$ of a 5nm EuS/4QL $(Bi_{0.22}Sb_{0.78})_2Te_3$ heterostructure exhibits a sharp peak at $V_g$ = -13V when the chemical potential is swept across the Dirac point, with a maximum of ~ 19.6kΩ due to the ambipolar carrier contributions (**Fig. 2a** and Supporting Materials). This indicates the conduction of the 4QL $(Bi_{0.22}Sb_{0.78})_2Te_3$ film is indeed dominated by the Dirac SSs. The corresponding current-voltage ($I_{sd}$ - $V_{sd}$) curves exhibit a linear relation throughout the shifting of the chemical potential which indicates the absence of activation energy for transport and is consistent with gapless Dirac SSs on the surface (**Fig. 2b**).

The in-plane magnetized 3D TI preserves the gapless character of the Dirac SSs while shifting the Dirac cone in momentum-space (**Fig. 1c**). To isolate the spin-related effects of the Dirac SSs, we applied in-plane magnetic fields in the range of ± 650 Oe (*x*-axis in **Fig. 1c**). **Figure 3a** summarizes the PMR and PHR of a 5nm EuS/4QL $(Bi_{0.22}Sb_{0.78})_2Te_3$ heterostructure with representative curves at several $V_g$s. The PMR and PHR are defined as PMR(%) = $\frac{R_{xx}(B)-\min(R_{xx}(B))}{\min(R_{xx}(B))} \times 100$ and PHR = $\frac{V_y(B)}{I_x}$. The PMR shows a butterfly-shaped hysteresis loop and can be tuned by the $V_g$ to a maximum amplitude of over 0.4%. Interestingly, the maximum of PMR occurs at $V_g$ = 0V (**Fig. 3b**), slightly different from the peak of the $R_{xx}$ at $V_g$ = -13V, which corresponds to the Dirac point (**Fig. 2a**). Since $V_g$ serves to filter out bulk conduction, this observation suggests that in addition to Dirac SSs, the bulk conduction may also contribute towards the PMR. We note that there is more than an order-of-magnitude enhancement of PMR in the 5nm EuS/ 4QL $(Bi_{0.22}Sb_{0.78})_2Te_3$ heterostructures with Dirac-SSs-dominated conduction observed here as compared to EuS/ 20QL $Bi_2Se_3$ heterostructures with bulk-carriers-dominated conduction (PMR~ 0.02%) [22,23], indicating the vital role of the SS in PMR. While obtaining appreciable hysteretic PMR changes in novel materials is of importance for future development of spintronic applications, it can arise from a variety of mechanisms such as AMR [24,25], domain-wall scattering [26] and spin-Hall magnetoresistance [7,27]. Furthermore, a prior experiment attributed the hysteretic PMR in EuS/TI heterostructures to magnetic domain-wall-trapped 1D conduction channels [22].

As compared with PMR, PHE measurements provide more insight into the underlying transport processes. The upper panel in **Figs. 3a** shows clear hysteresis loops, demonstrating the first observation of PHE in an exchange-coupled FMI/TI heterostructure. Interestingly, the PHE amplitude (*i.e.* PHR) can also be tuned by the $V_g$, with a maximum at the position of the Dirac point ($V_g$ = -13V) showing a direct correlation with the peak of $R_{xx}$. We note that while the observed PHR of several Ω is orders of magnitude larger than the PHR in 2D metallic ferromagnets (typically on the order of a few mΩ) [3,4,28,29], the Hall angle in both systems is comparable due to the large $R_{xx}$ in the TI/EuS system. In addition, we point out that the absence of symmetry outside the hysteretic PHR loop is attributed to noise and background reduction, which can play a role when modest resistance changes are recorded over a substantial sweep time.

To explore the physical origin of the large PHR signal in EuS/TI heterostructures, transport measurements were performed on another 5nm EuS/4QL $(Bi_{0.22}Sb_{0.78})_2Te_3$ sample with the measuring current aligned at three different angles $\Phi$ with respect to the external in-plane magnetic field $B$ (**Fig. 4a-4c**). The pronounced PHE signal around $\Phi = 0º$ shows a small decrease at $\Phi = 30º$ and is completely suppressed at $\Phi = 90º$ (**Fig. 4d**). This phenomenon is in stark contrast to the PHE observed in ferromagnets and semiconductors, where the suppression of the signal was observed both around $\Phi = 0º$ and $90º$ [2-6]. Similar angular dependence was also found in a 5QL $(Bi_{0.22}Sb_{0.78})_2Te_3$ film (Supplementary Material), with both 4 and 5 QL $(Bi_{0.22}Sb_{0.78})_2Te_3$ films being in the 3D TI regime with gapless Dirac SSs [13-15]. In addition, we note that the PMR doesn't depend strongly on the angle between $I$ and $B$ (the bottom panels of **Figs. 4a to 4c**), suggesting that AMR doesn't play a large role in our system and that the PHE arises from a different mechanism. The latter is also supported by the different $V_g$s at which maximum PMR ($V_g = 0V$) and maximum PHR ($V_g = -13V$) are attained (**Fig. 3b**). This is in contrast with conventional AMR and PHE which are both governed by the scaling of Eqs. (1) and (2) with $(\rho_{//} - \rho_{\perp})$. Specifically, the maximum PHR is correlated with the peak in $R_{xx}$ ($V_g = -13V$) corresponding to the Dirac point and suggesting the unusual PHE observed here is very likely related to the Dirac SSs of TI films. We note that the similar PMR values obtained in the measurements and the reproducibility of the data exclude the degradation of the exchange-induced magnetization of the EuS/TI heterostructure.

To shed more light on the role of the Dirac SSs in the magneto-transport of EuS/TI heterostructures, we reduced the thickness of the $(Bi_{0.22}Sb_{0.78})_2Te_3$ films. It is established both theoretically and experimentally that in the 2D limit of a 3D TI film ($< 4QLs$ in $Sb_2Te_3$ and $< 6QL$ in $Bi_2Se_3$), hybridization between the bottom and top Dirac SSs can occur, resulting in a hybridization gap [13-15,30-33]. The gap-opening in our 3QL TI heterostructures is confirmed by the much larger response to $V_g$, as shown in **Fig. 2c**. Additional support comes from the non-linear $I_{sd}$ - $V_{sd}$ characteristics near the charge neutral point (CNP) (**Fig. 2d**). This $I_{sd}$ - $V_{sd}$ non-linearity could be attributed to the presence of a gap [34,35] or shallow traps [36] in the narrow-gap semiconductors, both of which suggest the absence of a gapless Dirac SS. We note that the $I_{sd}$ - $V_{sd}$ curves of 3QL TI heterostructure are linear in the *n*- and *p*-doped regimes (**Fig. 2d**), thus excluding a poor-quality film as the source of the observed non-linearity.

The 5nm EuS/3QL $(Bi_{0.22}Sb_{0.78})_2Te_3$ heterostructure exhibited different magneto-transport properties (**Figs. 3c and 3d**). The PHR signal is observed only in the *p*-doped region and vanishes in the CNP region. The noticeable suppression of both the PMR and PHE around the CNP is consistent with a gap formation at the Dirac point. This emphasizes the role of the gapless Dirac SSs on magneto-transport. We note that in the *p*-doped region, the observed PHR signal in this heterostructure was also suppressed when the magnetic field was rotated from the *x*-direction ($\Phi = 0°$) to the *y*-direction ($\Phi = 90°$) (Supporting Materials). The similar angular dependence observed in the *p*-doped region of 3QL $(Bi_{0.22}Sb_{0.78})_2Te_3$ films is likely due to the contribution of the Rashba-type bands in the hybridized SSs and/or bulk valence bands [14].

Below we will explore possible existing mechanisms for PHE in relation with our two major findings: (1) the PHR shows a peak at the Dirac point, as revealed by the gate dependence; (2) the PHR is maximized when the in-plane magnetic field *B* is parallel to the current *I* ($\Phi = 0°$) but it is suppressed for *B* perpendicular to *I* ($\Phi = 90°$). It has been found that an out-of-plane ferromagnetic order exists at the interface between EuS and highly *n*-doped $Bi_2Se_3$ and can persist up to room temperature [37]. Out-of-plane magnetization can induce a large anomalous Hall (AH) response for chemical potentials in the vicinity of the CNP. However, no hysteresis loops or nonlinear features were observed in our Hall measurements on a 5nm EuS/4QL $(Bi_{0.22}Sb_{0.78})_2Te_3$ heterostructure with an out-of-plane magnetic field (Supporting Materials), thus excluding the possibility that the observed PHE comes from a weak out-of-plane ferromagnetism due to the misalignment of magnetic fields. In addition, the out-of-plane magnetoresistance did not exhibit the characteristic weak localization features for a magnetization-induced gap opening [38]. The emergence of a PHE was previously explained by a variety of mechanisms, including anisotropic scattering by magnetic impurities [2], a non-spherical Fermi surface [5], spin Hall magnetoresistance [7] and chiral anomaly [8-10]. However, these mechanisms satisfy the relationship described by Eqs. (1) and (2) and although they might play some role in our observations, they cannot explain the observed large PHR at $\Phi = 0º$ as well as the different scaling of the PHE and PMR. We note that deviations from Eq. (1) due to anisotropic resistivity in single crystals [39,40], such as in $SrRuO_3$ films [41-43], are also unlikely since the Hall bars used in our experiments were patterned by hand, and therefore repeated alignment with specific crystal axis is unlikely. Furthermore, no such deviations were found in high magnetic field measurements of TIs [6]. Additional deviations were attributed to strong magnetic anisotropy, similar to (Ga, Mn)As films [4] and $La_{1-x}Sr_xMnO_3$ films [40], but such an anisotropy is absent in the epitaxial EuS films [44]. Therefore, the unconventional PHE observed in $EuS/(Bi_{0.22}Sb_{0.78})_2Te_3$ heterostructures cannot be explained satisfactorily by the microscopic mechanisms discussed above.

To understand our observation of PHE, we next present a symmetry argument of the Hall response. The standard (linear) Hall response is described by $j_y = \sigma_{yx} E_x$, where $j_y$ is the Hall current, $E_x$ is the driving electric field and $\sigma_{yx}$ is the Hall conductivity. Without loss of generality, we may consider the magnetization of our system to be along the *x* direction ($\Phi = 0º$) which implies the symmetry of the system with respect to in-plane mirror operation $m_x$ ($x \rightarrow -x$, $y \rightarrow y$). As a result, the Hall conductivity $\sigma_{yx}$ must be zero since $E_x \rightarrow -E_x$, $J_y \rightarrow J_y$ under this symmetry operation. This symmetry argument is consistent with the vanishing PHR for $\Phi = 0º$ according to Eq. (1), and thus any linear Hall response mechanism cannot explain the observed non-zero PHE for $\Phi = 0º$ in our experiment. This motivates us to consider the non-linear Hall response, defined as $j_y = \sigma_{yxx} E_x^2$. Since $E_x^2$ is invariant under the mirror operation $m_x$, the above symmetry argument cannot exclude a non-zero non-linear Hall conductivity $\sigma_{yxx}$. Thus, our experimental observation should have non-linear Hall mechanism origin. Below, we will discuss two possible scenarios for the observed PHE in our experiment.

The first scenario is attributed to spin-orbit torque, which has previously demonstrated in a magnetic TI film with an in-plane magnetic field parallel to the current [45,46]. According to this scenario, a current in the *x*-direction will be accompanied by an effective magnetic field $B_{SO} = -I\lambda_{SO}\hat{\mathbf{y}} \times \mathbf{m}$, with $\mathbf{m}$ being the magnetization vector, $I$ is the current and $\lambda_{SO}$ is the spin-orbit coupling strength of surface states. When the external magnetic field is in parallel to the current (*x* direction), the effective magnetic field $B_{SO}$ is expected to possess an out-of-plane component, which can induce an out-of-plane magnetization component at equilibrium through the additional spin-orbit torque term $\tau_{SO} = -\gamma \mathbf{M} \times B_{SO}$ in the Landau-Lifshitz-Gilbert equation. The resulting out-of-plane magnetization can in turn give rise to an AH resistance, which is proportional to the current. Thus, this AH response induced by external in-plane magnetic fields is non-linear. On the other hand, when the external magnetic field is perpendicular to the current (*y* direction), the effective magnetic field $B_{SO}$ vanishes and no spin-orbit torque term appears. This is consistent with our observation that PHR is maximized for Φ = 0º but suppressed for Φ = 90º. Previous studies on magnetic TI films [45,46] have shown the spin-orbit torque term will maximize at the charge neutral point of top surface state, which agrees with our observation. However, we need to point out that while both surface states in magnetic TI films can contribute to spin-orbit torque [45,46], only the top surface state can couple to magnetic moments in EuS layer in our EuS/TI heterostructure. Therefore, to confirm this scenario additional experimental techniques which are beyond the scope of the present work, such as spin pumping [47-49], spin torque ferromagnetic resonance [50-52] and spin Seebeck effect [53] are required to directly probe spin torque in EuS/TI heterostructures.

The second scenario is related to the non-linear response of Dirac fermions to the in-plane magnetic field. We examine the effective Hamiltonian of the Dirac SS of TI with an in-plane magnetization [54],

$$H_0 = Dk^2 + \hbar v_f(k_x \sigma_y - k_y \sigma_x) + M \cos \Phi \, \sigma_x + M \sin \Phi \, \sigma_y \tag{3}$$

Here the kinetic energy term is expanded up to the second order in momentum ($Dk^2$) besides the linear term and the exchange-induced magnetization is described by $M$ and the angle $\Phi$. One can easily show that the Berry curvature for $H_0$ is always zero except at one gapless point $k_x = -\frac{M \sin \Phi}{\hbar v_f}$, $k_y = \frac{M \cos \Phi}{\hbar v_f}$, which indicates a vanishing linear Hall contribution. The non-linear Hall conductivity $\sigma_{yxx}$ can be evaluated through the perturbation theory and is given by

$$\sigma_{yxx} = \frac{3e^2}{2\pi} \sum_{k,\eta \neq \xi} \langle \phi_{k\eta} | \frac{\partial \phi_{k\xi}}{\partial k_y} \rangle \langle \phi_{k\xi} | \frac{\partial \phi_{k\eta}}{\partial k_x} \rangle \left( (J_x)_{\eta\eta}(k) - (J_x)_{\xi\xi}(k) \right) \frac{(\rho_{0,\xi\xi}(k) - \rho_{0,\eta\eta}(k))}{(E_{k\eta} - E_{k\xi})^2} \tag{4}$$

where $E_{k\eta(\xi)}$ and $\phi_{k\eta(\xi)}$ are eigen-energy and eigen-state with the index $\eta$ ($\xi$), $(J_x)_{\eta\eta} = \frac{1}{\hbar} \frac{\partial E_\eta}{\partial k_x}$ is the current operator and $\rho_{0,\eta\eta}(k)$ is the equilibrium distribution function (See Supporting Materials for a detailed derivation). For the Hamiltonian $H_0$, Eq. (4) suggests that the non-linear $\sigma_{yxx}$ mainly arises from the inter-band transition between the two branches of Dirac SSs with the same momentum but opposite velocities. Direct calculations in the clean limit (the relaxation time $T_2 \to \infty$) give rise to the non-linear $\sigma_{yxx} = \frac{3De^3 M}{4\hbar^2 v_f \varepsilon_F^2} \cos \Phi$, where $\varepsilon_F$ is the Fermi energy. The $\sigma_{yxx} \propto \frac{1}{\varepsilon_F^2}$ dependence is consistent with the observation of PHE enhancement around the Dirac point, as revealed by the gate dependence measurement. It should be pointed out that the divergence of $\sigma_{yxx}$ at $\varepsilon_F \to 0$ should be rounded off by disorder scattering. Furthermore, this non-linear Hall mechanism can also explain the absence of PHE in the CNP region of the 5nm EuS/3QL $(Bi_{0.22}Sb_{0.78})_2Te_3$ heterostructures, since the hybridization gap formed by quantum confinement in thinner TI film will lead to an enhancement of energy denominator in Eq. (4) and thus suppress the non-linear $\sigma_{yxx}$, in particular for chemical potentials near the CNP. Finally, although we cannot fully support the cosΦ dependence, the above analysis indicates that the non-linear $\sigma_{yxx}$ is maximized for Φ= 0º, but suppressed for Φ= 90º, in agreement with our observations (**Fig. 4d**). We note that the presence of a PHE signal in the *p*-doped region can be explained by the existence of Rashba-type bands in the hybridized SSs and/or bulk valence bands [14] since the relative position of the CNP is close to the maximum of the bulk valence bands [18,55,56]. In our experiment, the PHE is not observed in the *n*-type region of 3QL heterostructures. This is not surprising since the band structure of TI in the *n*- and *p*-type regions are usually not symmetric [14].

To summarize, our studies reveal the first observation of a significant PHE signal in FMI/TI heterostructures, demonstrating a unique dependence on TI film thickness, chemical potential and the angle between the in-plane magnetic field and the current. Based on a simple symmetry argument, we explain why linear Hall contributions cannot account for our observations and suggest two alternative non-linear contributions, with more experiments required to clarify the exact contribution of each mechanism. Our work will pave the way for the investigations of the topological magnetization dynamics and promote FMI/TI heterostructures as a platform for potential topological spintronic and electronic applications.

## Acknowledgments

The authors would like to thank N. Samarth and J. Shi for the helpful discussions. D.R, J.S.M, and C.Z.C. acknowledge support from grants NSF (DMR-1207469), ONR (N00014-13-1-0301), and the STC Center for Integrated Quantum Materials under NSF grant DMR-1231319. C.X.L acknowledges the support from the Office of Naval Research (Grant No. N00014-15-1-2675). C.Z.C. thanks the support from Alfred P. Sloan Research Fellowship and ARO Young Investigator Program Award (W911NF1810198).

**Figures:**

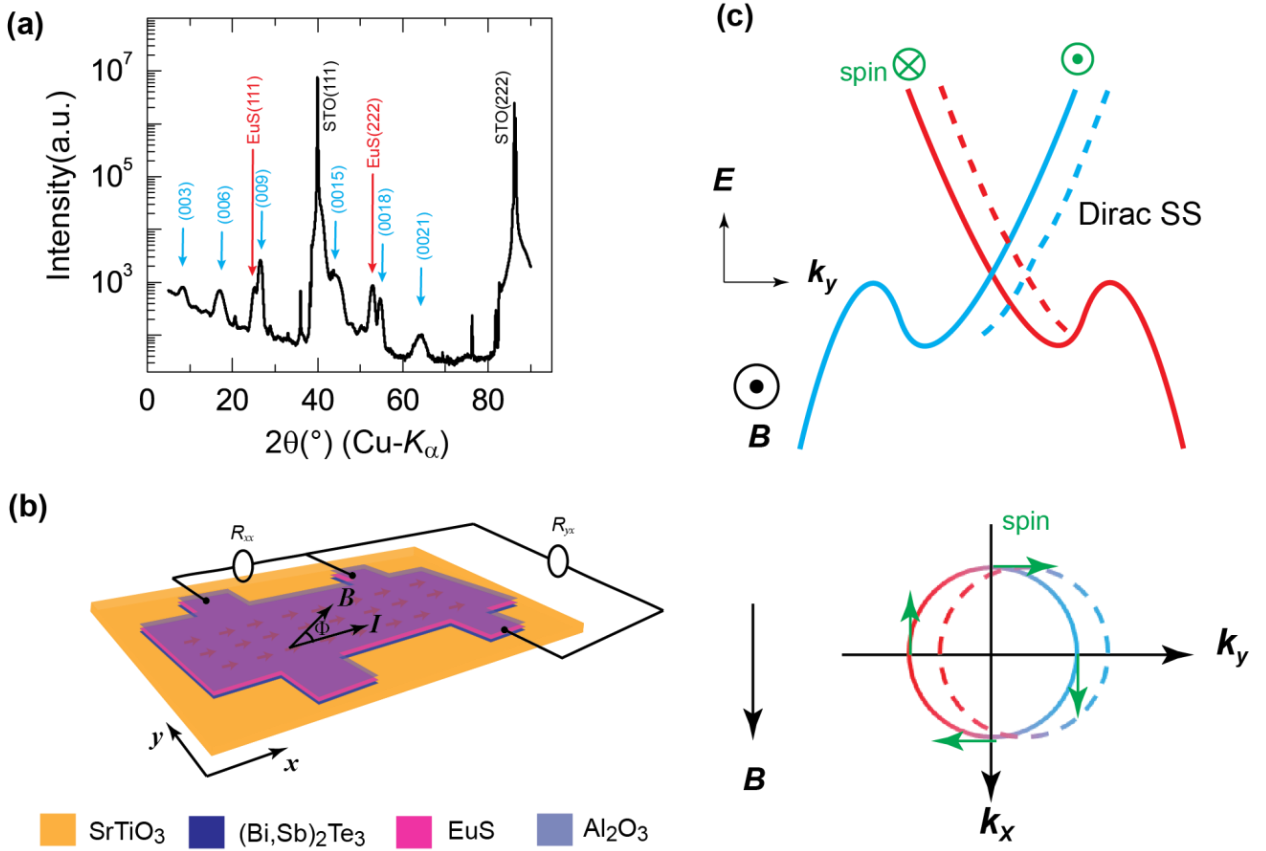

**Figure 1. The EuS/(Bi$_{0.22}$Sb$_{0.78}$)$_2$Te$_3$ heterostructures**. (a) X-ray diffraction spectrum of a 5nm EuS/4QL (Bi$_{0.22}$Sb$_{0.78}$)$_2$Te$_3$ heterostructure, with the peaks of the epitaxial (Bi$_{0.22}$Sb$_{0.78}$)$_2$Te$_3$ film and EuS film identified by blue and red arrows, respectively. (b) Schematic drawing of the sample structure and the Hall bar configuration. An image of the real Hall bar is shown in the Supporting Materials. (c) Schematic diagrams showing the shift of the Dirac SSs in the presence of the in-plane magnetization. The linear surface bands are a bit tilted due to the quadratic term $Dk^2$ included in the Hamiltonian of the Dirac surface states. The green arrows denote the spin of the Dirac electrons.

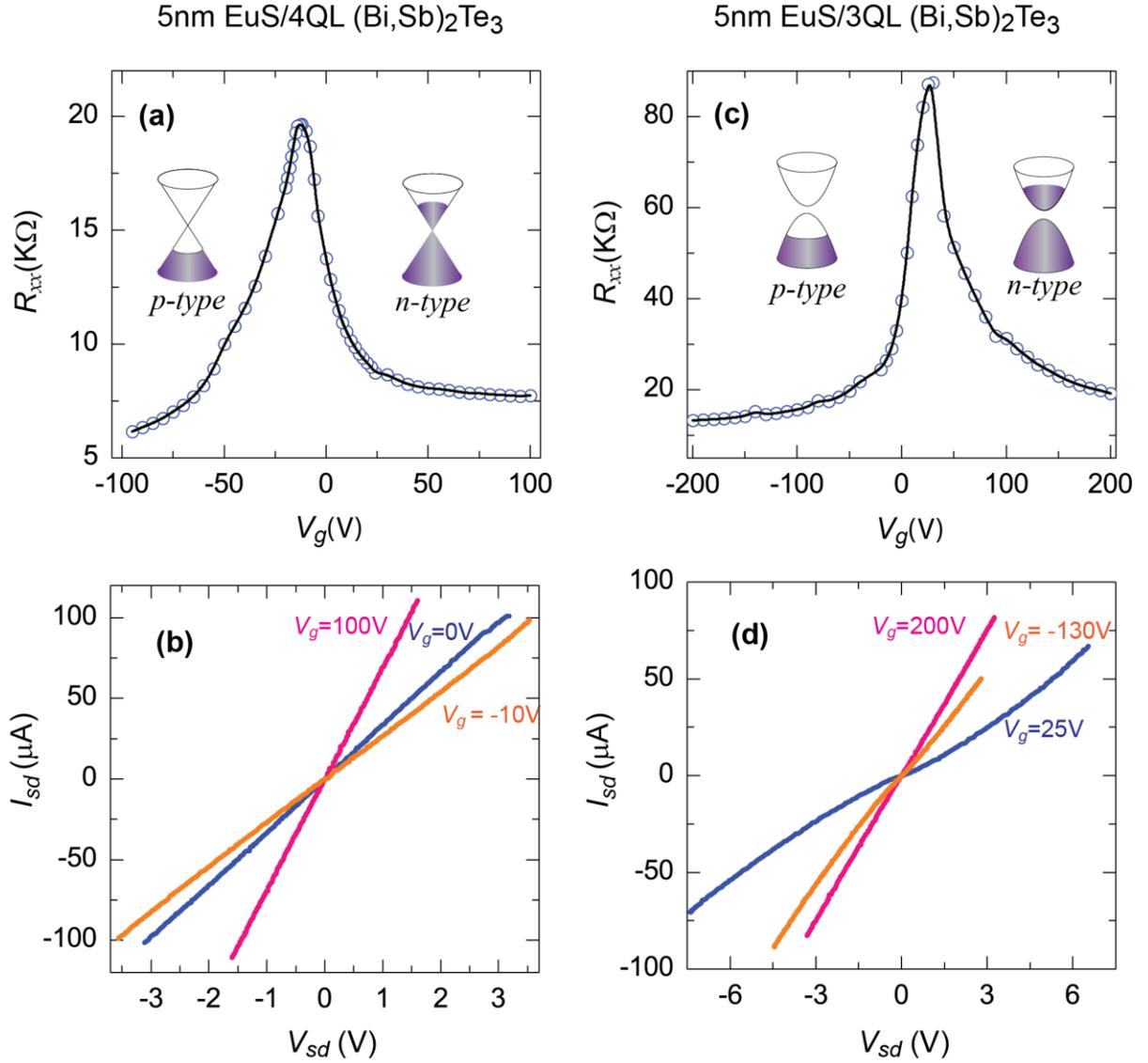

**Figure 2. Gate-dependent measurements on EuS/(Bi$_{0.22}$Sb$_{0.78}$)$_2$Te$_3$ heterostructures**. The gate ($V_g$) dependence of the sheet longitudinal sheet resistance ($R_{xx}$) of a 4QL (with Dirac SSs) and 3QL (with a hybridization gap) heterostructures is shown in (a) and (c), respectively. The corresponding $I_{sd}$-$V_{sd}$ curves for several $V_g$s are shown in (b) and (d).

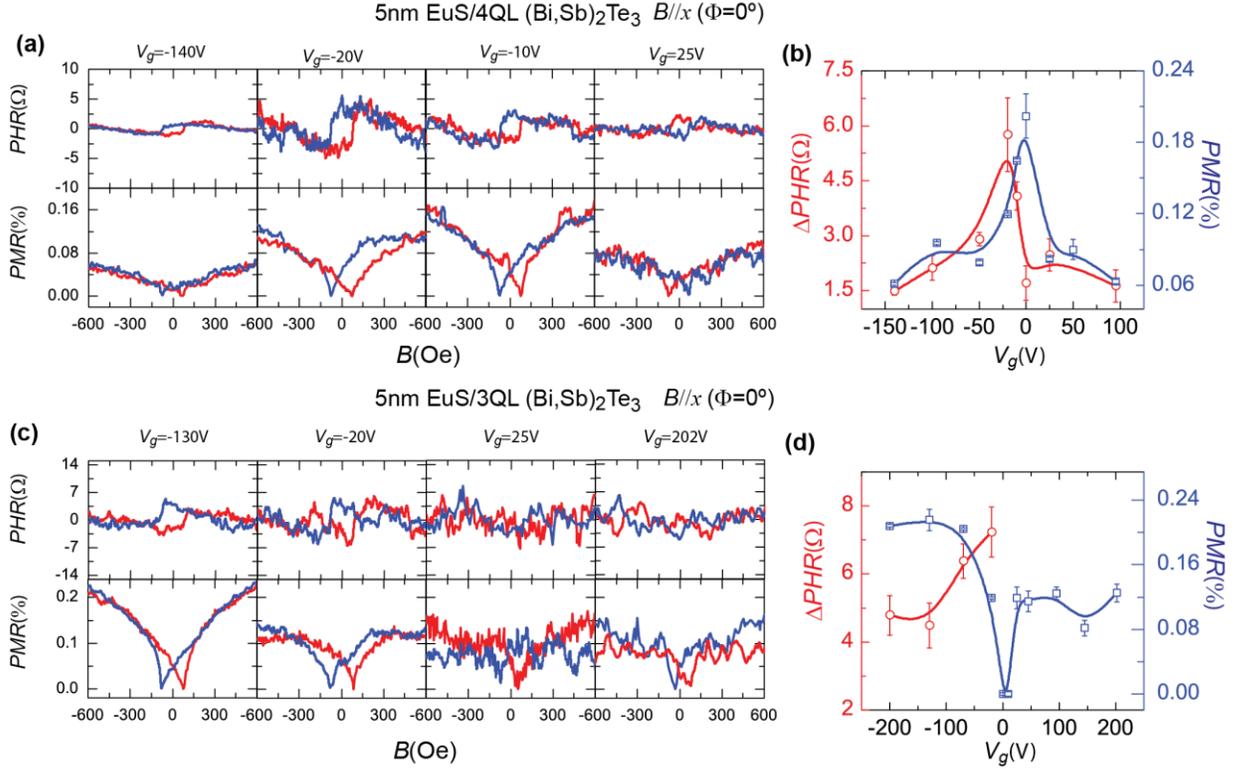

**Figure 3. In-plane magnetotransport in EuS/(Bi$_{0.22}$Sb$_{0.78}$)$_2$Te$_3$ heterostructures**. (a, c) Representative PHE and PMR measurements for several $V_g$s taken on 5nm EuS/4QL (Bi$_{0.22}$Sb$_{0.78}$)$_2$Te$_3$ (with Dirac SSs) (a) and 5nm EuS/3QL (Bi$_{0.22}$Sb$_{0.78}$)$_2$Te$_3$ (with a hybridization gap) (c) heterostructures, respectively. (b, d) The summary of the $V_g$ dependence of PHR and PMR for 5nm EuS/4QL (Bi$_{0.22}$Sb$_{0.78}$)$_2$Te$_3$ (b) and 5nm EuS/3QL (Bi$_{0.22}$Sb$_{0.78}$)$_2$Te$_3$ (d) heterostructures, respectively. The error bars for the PHR quantify the deviations in the signal around the limits of the transition, while the error bars in PMR quantify the deviations around 600 Oe. The zero PMR value in (d) was assigned due to the inability to observe a clear MR dependence while the error bar quantifies the fluctuations in the signal.

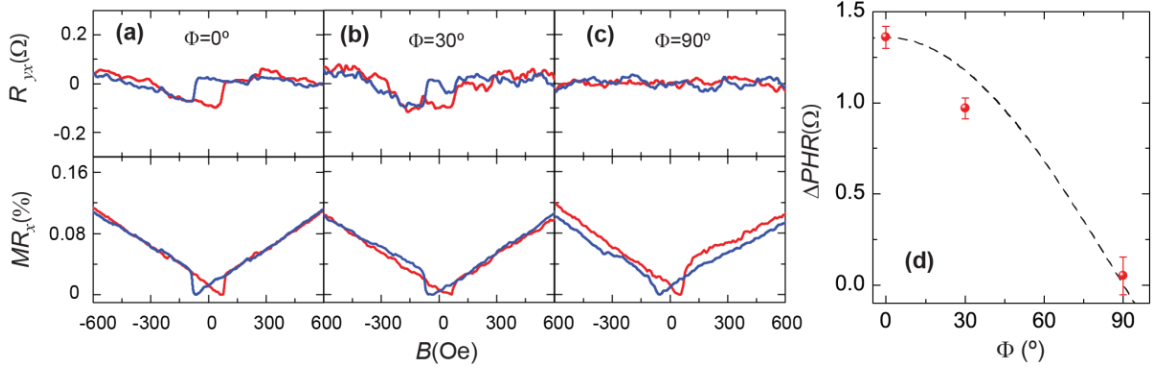

**Figure 4. Angle-dependent PHE and PMR measurements in another 5nm EuS/4QL (Bi$_{0.22}$Sb$_{0.78}$)$_2$Te$_3$ heterostructure.** (a-c) PHE (top panels) and PMR (bottom panels) taken on another 5nm EuS/4QL (Bi$_{0.22}$Sb$_{0.78}$)$_2$Te$_3$ (with Dirac SSs) heterostructure with Φ= 0º (a), Φ= 30º (b) and Φ= 90º (c), respectively. Φ is the angle between the current *I* and the in-plane magnetic field *B*. (d) Summary of the PHR as a function of Φ, the dashed line is a fit to A·cosΦ (A is the amplitude).